\documentclass[preprint2]{aastex}

\def\lesssim{\mathrel{\hbox{\rlap{\hbox{\lower4pt\hbox{$\sim$}}}\hbox{$<$}}}}
\def\gtrsim{\mathrel{\hbox{\rlap{\hbox{\lower4pt\hbox{$\sim$}}}\hbox{$>$}}}}
\newcommand{\deltavec}{\mbox{\boldmath $\delta$}}
\newcommand{\thetavec}{\mbox{\boldmath $\theta$}}
\newcommand{\zetavec}{\mbox{\boldmath $\zeta$}}
\newcommand{\xivec}{\mbox{\boldmath $\xi$}}
\newcommand{\etavec}{\mbox{\boldmath $\eta$}}

\newcommand{\zvec}{\mbox{\boldmath $z$}}

\begin{document}

\title{Signs of Planetary Microlensing Signals}

\author{Cheongho Han}
\affil{Department of Physics, Institute for Basic Science Research,
       Chungbuk National University, Chongju 361-763, Korea;
       cheongho@astroph.chungbuk.ac.kr}

\author{Kyongae Chang}
\affil{Department of Physics, Chongju University, Chongju 360-764, 
       Korea; kchang@chongju.ac.kr}

\begin{abstract}
An extrasolar planet can be detected via microlensing from the 
perturbation it makes in the smooth lensing light curve of the primary.  
In addition to the conventional photometric microlensing, astrometric 
observation of the center-of-light motion of the source star image 
provides a new channel of detecting and characterizing extrasolar 
planets.  It was known that the planet-induced astrometric signals 
tend to be positive while the photometric signals can be either 
positive or negative.  In this paper, we analytically show the 
reason for these tendencies of microlensing planetary signals.

\end{abstract}


\section{Introduction}
A microlensing event occurs when a lensing object approaches very 
close to the line of sight toward the background source star.  
Due to lensing, the lensed star appears to be split into two images.  
The locations and magnifications of the individual images are
\begin{equation}
\thetavec_\pm = {1\over 2}\left[ \zetavec \pm \sqrt{u^2+4}
{\zetavec\over \zeta} \right]\theta_{\rm E},
\end{equation}
and  
\begin{equation}
A_\pm = {\zeta^2+2\over 2\zeta \sqrt{\zeta^2+4}}\pm {1\over 2},
\end{equation}
where $\zetavec$ is the projected lens-source separation vector normalized 
by the Einstein ring radius $\theta_{\rm E}$.  The Einstein ring 
represents the effective lensing region around the lens within which 
the source star flux is magnified greater than $3/\sqrt{5}$.  
For a typical Galactic microlensing event, the Einstein ring radius is 
\begin{equation}
\theta_{\rm E} \sim 0.72\ {\rm mas}\ 
\left( {m\over 0.5 M_\odot} \right)^{1/2}
\left( {D_{os}\over 8\ {\rm kpc}}\right)^{-1/2}
\left( {D_{os}\over D_{ol}} -1\right)^{1/2},
\end{equation}
where $m$ is the lens mass and $D_{ol}$ and $D_{os}$ are the distances 
to the lens and source star, respectively.  For a rectilinear lens-source 
relative motion, the separation vector is related to the lensing parameters 
by
\begin{equation}
\zetavec = \left( {t-t_0\over t_{\rm E}}\right) \hat{\xivec} + 
\beta \hat{\etavec},
\end{equation}
where $t_{\rm E}$ is the time required for the source to transit 
$\theta_{\rm E}$ (Einstein time scale), $\beta$ is the closest lens-source 
separation normalized by $\theta_{\rm E}$ (impact parameter), $t_0$ is the 
time at that moment, and the unit vectors $\hat{\xivec}$ and $\hat{\etavec}$ 
are parallel with and perpendicular to the direction of the relative 
lens-source motion, respectively.  The image with the higher magnification 
$A_+$ (major image) is located outside of the Einstein ring, while the 
other image with the lower magnification $A_-$ (minor image) is located 
inside of the Einstein ring.  The separation between the two images, 
$\left\vert \thetavec_+ -\thetavec_-\right\vert=\sqrt{\zeta^2+4} 
\theta_{\rm E}$, for a typical Galactic event is very small and thus they 
cannot be resolved.  However, a lensing event can be identified from its 
characteristic smooth and symmetric light curve \citep{paczynski86}, which 
is represented by
\begin{equation}
A = A_+ + A_- = {\zeta^2+2\over \zeta \sqrt{\zeta^2+4}}.
\end{equation}
For a more detailed description about microlensing, see \citet{paczynski96}.

If a lensing event is caused by a star having a planet and the planet 
happens to locate close to the path of one of the two images produced 
by the primary star, the planet perturbs the nearby image and the event 
can exhibit noticeable deviations from the light curve of a single lens 
event \citep{mao91}.  It is empirically known that if the planet perturbs 
the major image, the resulting deviation in the lensing light curve 
becomes positive\footnote{For typical planet-induced perturbations, 
the perturbed part of the light curve is composed of one major peak and 
surrounding deviations with signs opposite to that of the peak deviation.  
Compared to the peak deviation, the deviations at the wings of the peak 
deviation are very small.  Throughout the paper, therefore, the sign of 
the planetary deviation implies that of the peak.}, while the deviation 
becomes negative if the planet perturbs the minor image 
\citep{gaudi97, wambsganss97, bozza99}.

For a planet with a mass ratio to the primary star of $q$, the planetary 
signal endures for a short period of time of $\sim\sqrt{q} t_{\rm E}$, 
corresponding to several days for a Jupiter-mass planet (with 
$q\sim {\cal O}10^{-3}$) and a few hours for an Earth-mass planet (with 
$q\sim {\cal O}10^{-5}$).  However, the strength of the signal depends 
weakly on $q$.  Then, planets can be detected if lensing events are 
monitored with a high enough frequency.  This frequency can be achieved 
from the combination of the early warning system to issue alerts of 
ongoing events in the early stage of lensing magnification and the 
follow-up observation program to intensively monitor the alerted events.  
Alert systems were and are being operated by the MACHO \citep{alcock96}, 
EROS \citep{afonso01}, MOA \citep{bond01} and OGLE \citep{udalski94} groups.  
The previously operated and currently operating follow-up collaborations 
include GMAN \citep{alcock97}, MPS \citep{rhie99}, PLANET \citep{albrow98}, 
and microFUN (D.\ DePoy, private communication).

As an additional channel to detect and characterize extrasolar planets via 
microlensing, \citet{safizadeh99} proposed astrometric follow-up observations 
of lensing events by using next generation high precision interferometers, 
such as those to be mounted on space-based platform, e.g.\ the {\it Space 
Interferometry Mission} (SIM), and those to be mounted on very large 
ground-based telescopes, e.g.\ Keck and VLT.  When an event is observed by 
using these instruments, although resolving the individual images is still
difficult, it is possible to measure the center-of-light motion of the 
lensed source star image caused by the change of the separation between 
the lens and source and the resulting variation of the brightness ratio 
between the two images.  For a single lens event, the shift of the image 
centroid with respect to the unlensed position of the source star (centroid 
shift) is represented by
\begin{equation}
\deltavec = {A_+\thetavec_+ + A_-\thetavec_-\over A} - 
\zetavec\theta_{\rm E} = 
{\zetavec \over \zeta^2+2}\theta_{\rm E}.
\end{equation}
The trajectory of the centroid motion (astrometric curve) traces out an 
ellipse during the event \citep{walker95, jeong99, dominik00}.  
\citet{safizadeh99} showed that planets can be identified from the 
perturbations in astrometric curves, which are analogous to photometric 
perturbations in lensing light curves.  They pointed out that due to the 
strong correlation between the photometric and astrometric planetary 
signals, adding astrometric information to the photometric lensing light 
curve will greatly help in determining the mass ratio and the projected 
separation of the planet.  In addition, since astrometric lensing 
observations enable one to determine the absolute mass of the lens system 
by measuring both the lens proper motion and parallax \citep{miyamoto95, 
hog95, walker95, paczynski98, boden98,gould99}, one can determine the 
absolute mass of the planet.  \citet{han02a} further investigated the 
patterns of astrometric deviations caused by planets with various 
separations and mass ratios.  From this investigation, they found an 
interesting tendency of astrometric planetary signals, where while 
photometric deviations can become either positive or negative depending 
on which of the two images produced by the primary is perturbed by the 
planet, astrometric deviations are positive in all tested events regardless 
of which image is perturbed.  The usefulness of this tendency was soon 
noticed by \citet{han02b}, who pointed out that the problematic photometric 
degeneracy between binary source and planetary perturbations \citep{gaudi98} 
can be unambiguously resolved with the additional astrometric information 
because the astrometric perturbations induced by a faint binary source 
companion are always negative, which is opposite to the sign of the 
planet-induced perturbations.  However, none of previous works explained 
the reason for the known tendencies of photometric and astrometric 
planetary signals.  In this paper, we analytically show why planet-induced 
astrometric signals are {\it always} positive while photometric signals 
can be either positive or negative.

The layout of the paper is as follows.  In \S\ 2, we describe the 
basics of planetary microlensing.  In \S\ 3, we derive the relation 
between photometric and astrometric planetary signals and explain 
the reasons for the empirically known properties of the signs of 
planetary signals.  We summarize and conclude in \S\ 4.

\section{Basics of Planetary Microlensing}

The planetary lensing behavior is described by the formalism of binary 
lensing with a very low mass-ratio companion.  If a source star located 
at $\zeta=\xi+i\eta$ in complex notations is lensed by two point-mass 
lenses with the individual locations of $z_{L,1}$ and $z_{L,2}$ and the 
mass fractions of $m_1$ and $m_2$, respectively, the locations of the 
resulting images $z=x+iy$ are obtained by solving the lens equation, 
which is represented by
\begin{equation}
\zeta = z + \sum_i {m_{i} \over \bar{z}_{L,i}-\bar{z}},
\end{equation}
where $\bar{z}$ denotes the complex conjugate of $z$ and all lengths 
are normalized by the combined Einstein ring radius.  Since the lens 
equation describes a mapping from the lens plane to the source plane, 
finding image positions ($x,y$) for a given source position ($\xi,\eta$) 
requires inverting the lens equation.  Although the lens equation for a 
binary lens system cannot be algebraically inverted due to its nonlinearity, 
it can be expressed as a fifth-order polynomial in $z$, and thus the 
image positions can be obtained by numerically solving the polynomial 
equation \citep{witt90}.  Since the lensing process conserves the source 
star surface brightness, the magnification of each image equals to the 
area ratio between the image and the unlensed source and mathematically 
it is given by the Jacobian of the mapping equation evaluated at the 
image position;
\begin{equation}
A_i = \left\vert \left( 1-{\partial\zeta\over\partial\bar{z}}
{\overline{\partial\zeta}\over\partial\bar{z}} \right)^{-1} \right\vert.
\end{equation}
Then, the total magnification is the sum of the magnifications of the 
individual images, i.e.\ $A=\sum_i A_i$.  Since the position of the 
image centroid equals to the magnification-weighted mean position of 
the individual images, the centroid shift is given by
\begin{equation}
\deltavec = {\sum_i A_i \zvec_i \over A} - \zetavec,
\end{equation}
where $\zvec_i$ and $\zetavec$ are the vector notations of each image 
position and the location of the unlensed source position, respectively.

Due to the very small mass ratio of the planet to the primary, the 
planetary lensing behavior is well described by that of a single lens 
for most of the event duration.  However, noticeable deviations can 
occur when the source passes the region close to caustics.  The caustics 
are the main new features of binary lensing and refer to the set of 
source positions at which the magnification of a point source becomes 
infinity.  For a planetary case, the caustics are located along or 
very close to the primary-planet axis ($x$ axis) and its location on 
the $x$ axis is approximated by 
\begin{equation}
x_c \sim x_p - {1\over x_p},
\end{equation}
where $x_p$ is the position of the planet \citep{griest98}.
Caustics 
are located within the Einstein ring when the planetary separation is 
in the range of $0.6\lesssim x_p\lesssim 1.6$.  Since the size of the 
caustic, which is directly proportional to the planet detection 
efficiency, is maximized when the planet is in this range, this range 
is referred as the `lensing zone' \citep{gould92}.  The location of the 
caustic in the {\it source} plane corresponds to the region near the one 
of the two images created by the primary in the {\it lens} (or {\it image}) 
plane.  From the lens plane point of view, therefore, noticeable deviation 
occurs when the planet is located close to one of the two images produced 
by the primary.

Major image perturbation refers to the case where the deviation is 
caused by a planet located near the major image.  Since the major 
image is located outside of the Einstein ring, major image perturbations 
are caused by planets with separations greater than $\theta_{\rm E}$, 
i.e.\ $\left\vert x_p \right\vert > 1$ (wide planet).  In this case, 
one finds from equation (10) that ${\rm sign}(x_c) ={\rm sign} (x_p)$ 
and $\left\vert x_c \right\vert<\left\vert x_p \right\vert$, implying 
that the caustic is located on the same side of the planet with respect 
to the center of mass between the primary and planet.

Minor image perturbation, on the other hand, refers to the case where 
the planet perturbs  the minor image.  The minor image is located 
inside of the Einstein ring, and thus minor image perturbations are 
caused by planets with separations less than $\theta_{\rm E}$, 
i.e.\ $\left\vert x_p \right\vert < 1$ (close planet).  Unlike the 
single caustic formed by the wide planet, the close planet causes 
formation of two caustics, which are located symmetrically with respect 
to the primary-planet axis, i.e. $x$ axis.  For planets in the lensing 
zone, however, the caustics are located very close to the $x$ axis, and 
thus their locations can also be approximated by eq.\ (10).  Since 
${\rm sign}(x_c) \neq {\rm sign}(x_p)$, the caustics are located on the 
opposite side of the planet with respect to the center of mass.

\section{Signs of Planetary Perturbations}

In this section, we analytically derive the relation between photometric 
and astrometric microlensing signals of planets and explain the reason 
for the empirically known properties of the signs of the signals.

We begin with the case where the major image is perturbed by a planet.  
Let us define $\epsilon$ as the fractional photometric deviation, i.e.
\begin{equation}
\epsilon = {A_p-A\over A},
\end{equation}  
where $A_p$ and $A=A_+ + A_-$ represent the magnifications with and 
without the perturbation, respectively.  Since the minor image is not 
perturbed by the planet (see Appendix), the perturbed magnification 
is $A_p = A_{p+} + A_-$, where $A_{p+}$ represents the magnification 
of the perturbed major image.  Then, the magnification excess can be 
written as 
\begin{equation}
\epsilon = {A_{p+}-A_+\over A} = {A_{p+}-(A-A_-) \over A}.
\end{equation} 
By inverting eq.\ (12), the perturbed major image magnification is 
expressed in terms of $\epsilon$ by
\begin{equation}
A_{p+} = (1+\epsilon)A - A_-
\end{equation}

The location of the perturbed image centroid is represented by
\begin{equation}
\varphi_p = {A_{p+}\theta_+ +A_-\theta_- \over A_{p+}+A_-},
\end{equation}
where $\theta_+$ and $\theta_-$ are the positions of the unperturbed 
major and minor images, respectively.  Here we use an approximation 
that the change in the position of the major image due to the planetary 
perturbation is small.\footnote{When an images produced by the primary 
is perturbed by the planet, the fractional astrometric shift of the 
perturbed image is 
$$
{\Delta x\over x} \sim {q\over x-x_p} {1\over \zeta}
\propto (x-x_p)^{-1},
$$
where $x$ is the location of the image, $x_p$ is the location of the 
planet, and $\zeta$ is the primary-source separation [see equations 
(35) and (36) of \citet{bozza99}].
On the other hand, the fractional photometric deviation is 
$$
{\delta A\over A} \sim 
{2q\over (x-x_p)^2} {1\over x^2-1/x^2} \propto 
(x-x_p)^{-2}
$$
(see Appendix).
During the time of maximum perturbation when $x_p\rightarrow x$, 
therefore, the photometric perturbation dominates over the astrometric 
perturbation, i.e.
$\delta A/A \gg \delta x/x$.
This implies that the astrometric perturbation to the centroid shift 
derives largely from the photometric perturbation rather than the 
astrometric shifts in the positions of the individual images.
}
By plugging eq.\ (13) into eq.\ (14), the centroid position is 
expressed in terms of $\epsilon$ by
\begin{equation}
\varphi_p = {
(1+\epsilon)A\theta_+ + A_-(\theta_--\theta_+)
\over 
(1+\epsilon)A}.
\end{equation}

By definition, the astrometric planetary signal is the difference 
between the centroid positions with and without the planetary 
perturbation, i.e.
\begin{equation}
\Delta\varphi = \varphi_p - \varphi,
\end{equation}
where $\varphi = (A_+\theta_+ + A_-\theta_-)/A$.  From eq.\ (15) and 
(16), one finds that the relation between the astrometric and photometric 
perturbations in the case of the major image perturbation is
\begin{equation}
\Delta\varphi = {\epsilon\over 1+\epsilon} {A_-\over A} 
(\theta_+ - \theta_-),~~({\rm major~image~perturbation}).
\end{equation}
One finds a similar relation in the case of the minor image perturbation; 
\begin{equation}
\Delta\varphi = -{\epsilon\over 1+\epsilon} {A_+\over A} 
(\theta_+ - \theta_-),~~({\rm minor~image~perturbation}).
\end{equation}

If we define the signs of astrometric shifts such that the direction 
toward the major image from the lens position is positive, the sign of 
the term $\theta_+-\theta_-$ is also positive.  Magnification is always 
positive by definition, and thus the signs of both terms $A_-/A$ and 
$A_+/A$ are also positive.  Then, one finds the relation between the 
signs of photometric and astrometric perturbations in the case of the 
major image perturbation;
\begin{equation}
\cases{
\Delta\varphi_p > 0 & if $\epsilon > 0$, \cr 
\Delta\varphi_p < 0 & if $-1 <\epsilon < 0$. \cr
}
\end{equation}
In the case of the minor image perturbation, the 
relation is 
\begin{equation}
\cases{
\Delta\varphi_p < 0 & if $\epsilon > 0$, \cr 
\Delta\varphi_p > 0 & if $-1 <\epsilon < 0$. \cr
}
\end{equation}
As shown in Appendix, the flux of the major image is magnified 
($\epsilon > 0$) when it is perturbed by the planet, while the flux of 
the minor image is demagnified ($-1 <\epsilon < 0$) due to planetary 
perturbations.  Then, the sign of the astrometric perturbation is 
positive in both cases of major and minor image perturbations.
{\it Therefore, while the sign of the photometric planetary signal is 
either positive or negative depending on whether the major or minor is 
perturbed, astrometric signals are always positive regardless of the 
type of  image perturbations}.

In Figure 1, we present the geometry of an example planetary lens 
system where the major image is perturbed by a planet.  In the figure, 
the coordinates are centered at the center of mass of the lens 
system and the positions of the primary and the planet are marked by 
`x'.  The diamond-shaped figure represents the caustic and the big 
dashed circle is the combined Einstein ring.  The straight line with 
an arrow represents the source trajectory and the solid and dotted 
curves running almost in parallel with the source trajectory are the 
trajectories of the image centroid (with respect to the lens positions) 
with and without the planet-induced perturbation, respectively.  Also 
marked are the location of the source (the small solid circle on the 
source trajectory) at the moment of the image perturbation and the 
images (elongated figures at around $\xi=-1.4$ and 0.7) corresponding 
to the source position.  The blowup of the region around the major 
image is shown in the inset, where the perturbed and unperturbed 
images are drawn by solid and dotted lines, respectively.  From the 
comparison of the sizes of the unperturbed and perturbed images, one 
finds that the major image is {\it magnified} due to the perturbation.  
Photometrically, this causes positive deviations in the lensing light 
curve (lower right panel).  Astrometrically, due to the magnified flux 
of the major image, the image centroid is additionally shifted further 
away from the unlensed source position towards the major image, causing 
also positive deviations in the astrometric curve (lower left panel).  
Figure 2 shows several more examples of perturbed and unperturbed images 
to illustrate the mentioned trend of image perturbation applies to 
general cases of perturbations.

In Figure 3, we present the geometry of an example planetary lens system 
where the minor image is perturbed by a planet.  From the comparison of 
the unperturbed and perturbed minor images, one finds that the perturbed 
image is {\it demagnified}, in contrast with the magnified major image 
due to the planetary perturbation.  Astrometrically, demagnification of 
the minor image causes the image centroid to be further shifted towards 
the major image, resulting in positive deviations, whose sign is same as 
that of the astrometric deviations caused by the major image perturbation.

\section{Conclusion} 
We investigated the properties of planetary signals in microlensing 
light curves and centroid shift trajectories.  We derived analytic 
relation between the photometric and astrometric planetary signals 
and explained the reason for the empirically known properties of 
planetary signals, where the photometric signal can be either positive 
or negative depending on which of the two images produced by the primary 
is perturbed, while the astrometric signal is always positive regardless 
of which image is perturbed.

\acknowledgments
We thank B.\ S.\ Gaudi and V.\ Bozza for making very helpful comments on 
planetary microlensing properties.  CH was supported by the Astrophysical 
Research Center for the Structure and Evolution of the Cosmos (ARCSEC) 
of Korea Science \& Engineering Foundation (KOSEF) through Science Research 
Program (SRC) program.  KC was supported by Korea Astronomy Society (KAO).

\clearpage
\appendix

\centerline{\Large Appendix: Signs of Photometric Perturbations}
\bigskip

By treating the planet-induced deviation as a perturbation, \citet{bozza99} 
derived the expression for microlensing magnification of a lens with a 
planet by expanding the Jacobian of the lens equation to the first order 
in $q$ [see eq.\ (37) of his paper]; 
\begin{equation}
A_p = A + {A\over z^4-1}
\left\{ 
{2q(x^2-y^2)[(x-x_p)^2-y^2] + 4(x-x_p)^2y^2
\over 
[(x-x_p)^2+y^2]^2}
-
{{4(x\Delta x + y\Delta y)
\over z^2}
}
\right\},
\end{equation}
where $A$ is the unperturbed magnification, $\zvec=(x,y)$ is the position 
vector to the image location, $x$ axis is parallel to the primary-planet 
axis, $x_p$ is the planet position on the axis, and $(\Delta x, \Delta y)$ 
represents the change of the image position induced by the planet.  The 
perturbation becomes maximum when the source crosses $x$ axis.  At this 
moment, the images are also located along or very close to $x$ axis.  
Then, by using the approximations of $y\rightarrow 0$, 
$\Delta y\rightarrow 0$, and $z\rightarrow x$, one can express the 
equation into a one-dimensional form of 
\begin{equation}
A_p =  A \left\{
1+ {1\over x^2-1/x^2}
\left[ {2q\over (x-x_p)^2} -{4\over x^2}\left( {\Delta x\over x}
\right)\right] \right\}.
\end{equation}
As noted by \citet{bozza99}, the planet induces two types of perturbations: 
the first type caused by the slight change of the image position (the term 
$\propto \Delta x/x$) and the second type resulting from the change of the 
lens equation due to the planet [the term $\propto (x-x_p)^{-2}$].  
The fractional astrometric perturbation is 
$\delta x/x \propto (x-x_p)^{-1}$ \citep{bozza99}.
Then during the time of maximum perturbation when $x\rightarrow x_p$, 
the second type perturbation dominates and the magnification 
excess is approximated by
\begin{equation}
\epsilon = {A_p-A\over A} 
\sim {2q\over (x-x_p)^2} {1\over (x^2-1/x^2)}.
\end{equation}
From eq.\ (3), one finds that the planet's approach close to one image 
(located at $x_1$) causes little perturbation on the other image (located 
at $x_2$) because $(x_1-x_p)^2 \ll (x_2-x_p)^2$, implying that only 
one image is perturbed by the planet.  In addition, since both $q$ and 
the term $(x-x_p)^2$ are positive, the sign of the photometric 
perturbation is determined by the remaining term $x^2-1/x^2$.  Because the 
major image is located outside of the Einstein ring ($x > 1$), and vice 
versa in the case of the minor image perturbation, one finds that 
\begin{equation}
{\rm sign}\left( x^2- {1\over x^2} \right) = 
\cases{
(+),  & for major image perturbation,  \cr
(-),   & for minor image perturbation.  \cr }
\end{equation}
Therefore, the flux of the major image is magnified when it is perturbed 
by the planet, while the flux of the minor image is demagnified due to 
planetary perturbations, i.e.
\begin{equation}
{\rm sign}(\epsilon) = 
\cases{
(+),  & for major image perturbation, \cr
(-),   & for minor image perturbation.  \cr }
\end{equation}

\begin{figure}
\epsscale{1.3}
\centerline{\plotone{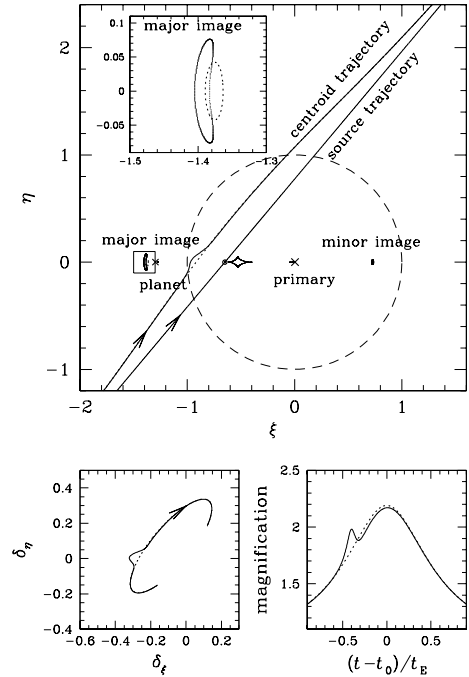}}
\vskip-0.6cm
\caption{
Upper panel: 
Geometry of a lens system where the major image is perturbed by a planet.  
The coordinates are centered at the center of mass of the lens system 
and the positions of the primary and the planet are marked by `x'.  The 
diamond-shaped figure represents the caustic and the big dashed circle 
is the combined Einstein ring.  The straight line with an arrow represents 
the source trajectory.  The small solid circle on the source trajectory 
represents the source position at the moment of the major image 
perturbation and the elongated figures at around $\xi=-1.4$ and 0.7 
represent the images corresponding the source position.  The inset shows 
the blowup of the region around the major image.  The solid and dotted 
curves running in almost parallel with the source trajectory are the 
trajectories of the image centroid with and without the planet-induced 
perturbation, respectively.  The mass ratio and the separation (normalized 
by $\theta_{\rm E}$) of the planet are $q=3\times 10^{-3}$ and $a=1.3$, 
respectively, and the source star has an angular radius of 
$0.02 \theta_{\rm E}$.
Lower panels:
The astrometric (left panel) and light curves (right panel) with (solid 
curves) and without (dotted curves) the planetary perturbation, respectively.
}\end{figure}

\begin{figure}
\epsscale{1.3}
\centerline{\plotone{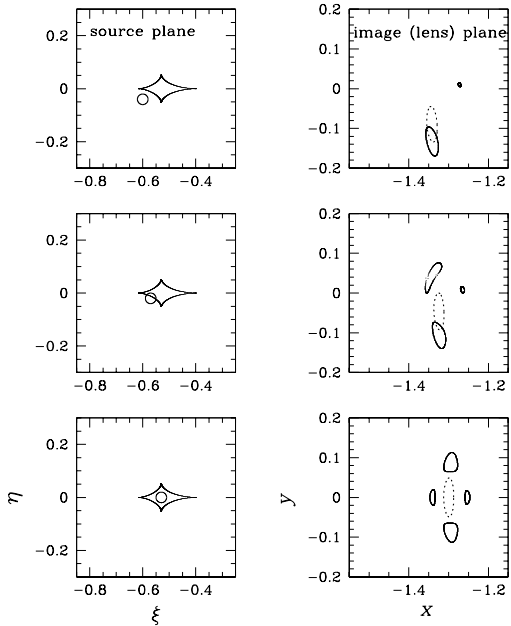}}
\vskip-0.6cm
\caption{
Variation of the image shape perturbed by a planet depending on the 
source position with respect to the caustic.  Left panels show the 
locations of the source star (small solid circle) with respect to the 
caustic (diamond-shaped figure) and right panels show the resulting 
images (closed figures drawn by solid curves) corresponding to the 
source positions.  The figures drawn by dotted curve in the right panel
are the unperturbed images.  The lens system is the same as in Fig.\ 1 
and the the planet perturbs the major image.
}\end{figure}

\begin{figure}
\epsscale{1.3}
\centerline{\plotone{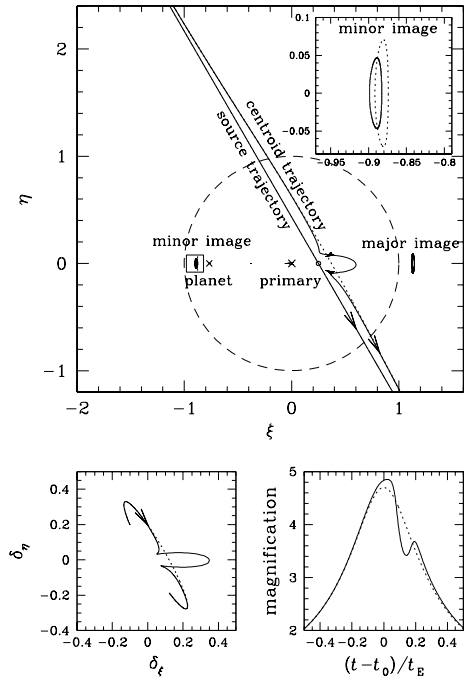}}
\vskip-0.6cm
\caption{
Geometry of a lens system where the minor image is perturbed by a planet.  
Notations are same those as those in Fig.\ 1.  Lens and source parameters 
are same as those of the lens system in Fig.\ 1 except that the planet 
separation is $a=1/1.3\sim 0.77$ and the source follows a different 
trajectory.
}\end{figure}

\end{document}